\documentclass[a4paper,11pt]{article}
\usepackage{pos}

\title{Neutrinos: Opening new doors}

\author{Francesco Vissani}

\affiliation{INFN, Laboratori Nazionali del Gran Sasso,\\
  Assergi (AQ), Via Acitelli 22, Italy}

\emailAdd{vissani@lngs.infn.it}

\abstract{In this contribution, which introduces the ‘Crossing the Portal’ session of the NOW 2024 meeting, I discuss the value of the concepts of interdisciplinarity and innovation for neutrino physics. 
After some historical considerations, which provide an initial illustration of the significant role of these concepts, I review some well-known cases of neutrino science, involving both astrophysics and particle physics, which allow us to deepen the analysis. 
The importance of a harmonious relationship between theoretical elaborations and experiments emerges: effective collaboration between theoretical and experimental physicists played a key role in many of the successful cases and proved marginal or defective in the doubtful ones.
The discussion highlights also the need to proceed armed with a patience, a virtue that Feynman himself indicated as essential to science~\cite{dick}, all the more necessary when investigating interesting but elusive particles such as neutrinos.

}
\FullConference{12th Neutrino Oscillation Workshop (NOW2024)\\
 2-8, September 2024\\
Otranto, Lecce, Italy\\}

\begin{document}
\maketitle

\section{Introduction}
Neutrino science is approaching its centenary and thus seems to be moving towards a stage of maturity.
Yet, even thinking only of the previous main conference, Nu'24 in Milan, we heard:  
of very high energy events by cosmic neutrinos;
of possible hints of supernovae throughout the history of the universe;
of strong new limits on neutrino mass, from the laboratory and cosmology;
of significant progress in the search for Majorana's mass...

In short, this {\em interdisciplinary} science continues to produce results and promise, i.e., {\em innovation}. As an introductory contribution to the session ‘Crossing the portal’ my proposal is to discuss these concepts with one question in mind: Is everything okay as it is or are we in danger of missing something important?

Before I begin, I would add a further call for caution, 
 as the two main terms in the debate have some room for ambiguity;
to highlight this, here is 
a list of words that we contrast with them:
\\ {\small
\indent $\bullet$  {\em for interdisciplinary}:
incomplete, exclusive, narrow, standard, specific, precise, fundamental;\\
\indent $\bullet$   {\em for innovation}:
stagnation, repetition, conventionalism, orthodoxy, preservation,  tradition.\\
} 
Their meanings range from `short-sighted' to `essential'  in the former case 
and from `obstructive' to `reliable' in the latter.  This 
 should put us on notice: seeking the good is more difficult than following a flag, whatever it may be. Or more simply, we must be careful to gain and not to lose.  

That said, let me present the exposure plan: 
I begin with some notes on the very early history of the neutrino; 
I continue with certain successful cases from astrophysics and particle physics; 
then I propose a few doubtful cases for consideration; 
finally, I conclude with brief notes and assessments.

\section{Notes from early history}
The innovative potential of neutrinos emerges from the fact that they have played a pivotal role in nuclear and  particle physics.
 We can back this up by considering a few early episodes \cite{rr}:\\ {\small
{\sf $\star$ Pauli 1930:} non-relativistic model of the nucleus with a ghost-like particle of matter in it;\\
{\sf $\star$ Perrin 1933:} 
neutrino emission can be better thought of as a radiated wave, just as the photon, in line with Ambarzumian \& Iwanenko 1930, who talk of the electron\footnote{The description / understanding of $\beta$-rays emission became particularly urgent in 1932, after neutron discovery and the 
novel model of the nucleus (Iwanenko, Heisenberg and Majorana) with neutrons and protons only.};\\
{\sf$\star$  Fermi 1933:} relativistic model of $\beta$ decay with old quantisation procedure, where $\nu\neq \bar\nu$ necessarily\footnote{I am not so sure that the traditional question whether the neutrino mass is `Dirac or Majorana' is phrased in the most appropriate terms, as I do not know of any paper in which Dirac talks about neutrinos (while Fermi surely did).};\\
{\sf $\star$ Majorana 1937:} devises QFT for fermions and points out the possibility $\nu= \bar\nu$,  resembling the photon.

}
\indent The importance of neutrinos for interdisciplinary investigations emerges soon after in relation to astrophysics and cosmology. It is enough to recall some of the early landmark works.
\\ {\small
{\sf $\star$  Bethe '30 + Peierls:} neutrinos  are {\em unobservable} (1934) and non-essential for solar energy-loss (1939);\\
{\sf $\star$  Gamow et al '40}: neutrinos are important for supernovas (1940) and cosmos (1948);\\
{\sf $\star$  Pontecorvo '46:} idea to measure solar neutrinos, even if those known at the time all have small interactions;\\
{\sf $\star$  Fowler et al '58:} predict that the PP chain branches gives rise also to $^7$Be and $^8$B neutrinos, better detectable\footnote{When it was realised that PP chain is dominant? This is given for granted in B$^2$FH paper (1957); Bahcall says that this was the general belief in '50. 
I suppose one should emphasise the key remark by Payne 1925, and compare with Bethe 1939, who 
assuming H=35\%, N=10\%  is led to believe that CNO dominant in the sun.}.\\
\indent 
(I have tried to illustrate the importance of neutrinos for the discussion of fundamental interactions, but it has resulted in such a complicated map that I refrain from presenting it. Just for fun, I will point out one interesting and little-known episode: the first speculations that neutral current weak interactions could exist dates back to two independent works by Gamow \& Teller and Kemmer written in 1937.)

Let's take now a big leap forward in time and move on to reflect on recent cases, focusing on the period when the standard model was finally becoming established.

\section{Cases from particle physics}
In this section I present  some  successful investigations that concern the neutrino mass question.
\paragraph{\em Neutrino masses in gauge theories:}
Initially, the interest in extended gauge theories  focused on proton decay. 
Peter Minkowki, one of the pioneers of SO(10), first spoke of neutrino mass in this context\footnote{He considered a left-right gauge theory.
As is well known, similar arguments were later explored and extended further by Yanagida; Glashow; Gell-Mann, Ramond and Slansky; Mohapatra and Senjanovi\'c.} (1977).
The smallness of the neutrino mass is attributed to a new scale of physics as ZZ Xing argued yesterday. This leads us to an open line of research
which includes experimental tests, such as $0\nu 2\beta$ (see below).
\paragraph{\em Neutrino propagation in  matter:} The theory of vacuum oscillations, which dates back to the 1960s, has been significantly extended, mainly thanks to Wolfenstein (1978), Mikheyev and Smirnov (1985-'86), 
who clarified the effect of propagation in an external medium.
This has allowed the development of methods for exploring the kind of mass spectra that have not yet exhausted their scientific potential.
\paragraph{\em The thirtieth anniversary of global analysis of neutrino masses:} In 1994, three of the organizers of NOW 2024 (including the session chairman G~Fogli) started a global program of analysis of neutrino oscillation data, which has borne much fruit over time. In the course of the years, other excellent working teams have joined the enterprise: today we can count on the professional data analyses of the Bari group, the Valencia group and NuFit. 
In this connection, let 
me stress that{\small
\begin{itemize}
\item 3 flavor analyses have always displayed consistency;\\[-4.2ex]
\item they are still crucial for the field after so many years;\\[-4.2ex]
\item they will continue to be so after JUNO, HyperK, DUNE.
\end{itemize}}
\noindent Finally, let me emphasize that - despite certain superficial impressions - this type of enterprise is configured as real theoretical physics, ostensibly not too ambitious, but progressive and cumulative. 
A revolutionary, unexpected and ultimately welcome discovery can sometimes come; but one must always remember that not infrequently, especially in neutrino physics, the only possible progress is one that happens gradually.
\paragraph{\em Absolute neutrino mass:} 
The very strong limits from cosmology, now hotly debated, have been consolidating for 10 years now, and  I would not forget Seljak et al.~'04. The recent DESI results show, in my opinion, that the method has great potential: it seems more and more reasonable to believe that the type of neutrino mass spectrum will be significantly and precisely probed thanks to cosmology. 
However, it is advisable to exercise prudence: In the past, some cosmologists have argued for non-zero values that we no longer believe in today
(e.g., Primack '94; Allen '03; Battye '14). Also laboratory physics had its premature claims, as the 30 eV neutrino (Lubimov ’80) or the 17 keV one (Simpson ’85), just to remind a few particularly well-known. On the contrary, I believe that the perseverance of KATRIN, which after 20 years is reaching its goal, is highly commendable, even if their sensitivity is less than that achieved by cosmological measurements.
\paragraph{\em Nature of the neutrino and Majorana theory:} 
Curiously the best chance to probe lepton number of the standard model is a nuclear physics process discussed almost 100 year ago:
neutrinoless double beta decay (Furry 1939) usually denoted by $0\nu 2\beta$.
The most plausible contribution to this process is due to Majorana's neutrino masses, see Greuling \& Whitten 1960.
This is bound (but not determined) by oscillations and other available data, as discussed in 
many works since Nu'98 in Takayama, beginning with mine  ('99).
The plausibility of $0\nu 2\beta$ lies in the very structure of the SU(2)$_{\mbox{\tiny L}}\times$U(1) electroweak theory. 
The value of the Majorana neutrino mass is the main reason why the rate of the transition is uncertain; if theory, cosmology or anything else could help to improve its knowledge, that would be great.

\section{Cases from astrophysics}
Here are some selected interdisciplinary   results concerning neutrinos.
\paragraph{\em Beginning of neutrino astronomy:}
In the mid-1960s, Ray Davis Jr and John Bahcall undertook a heroic adventure in nuclear, particle and astrophysical physics, with experimental, observational and theoretical aspects. After the discovery of the solar neutrino deficit, reservations were expressed mainly by the particle physics community; no further tests were conducted for 20 years. The theoretical model of Bahcall, however, was correct and provided (and continues to provide) valuable guidance and inspiration on how to proceed. E.g., H Chen's paper proposing the SNO experiment (1985) begins as follows:
\begin{quote}
\footnotesize
The solar-neutrino problem, i.e., fewer neutrinos
are assigned to the sun in the chlorine-argon radiochemical experiment of Davis and co-workers than
predicted by the standard solar model [...]
\end{quote}


\paragraph{\em  First tests and progresses:}
The leader of KamiokaNDE, Koshiba, was a CR physicist who later worked in accelerators. He met theorists as Yukawa, Nambu, Feynman, Fermi, Sato, Arafune, Occhialini, Sugawara; the latter ones had a special influence on him.  
%
The KamiokaNDE experiment was based on the photo-multipliers he developed. 
Started in '83 and originally motivated by extended gauge theories (NDE=nucleon decay experiment) KamiokaNDE enabled neutrino detection from SN1987A and the study of atmospheric neutrinos. The latter provided an additional clue to neutrino oscillation; their investigation at Super-KamiokaNDE (NDE=neutrino detector experiment), in operation since '96, allowed for definitive proof.

\paragraph{\em Solar neutrino spectroscopy:}
The measurement  of PP chain and CNO cycle neutrinos have been extraordinary results, enabled by the care with which the Borexino detector was prepared.  Much credit is due to the leader G~Bellini and the Borexino staff, including A~Ianni, N~Rossi, O~Smirnov, etc. 
A contribution to CNO detection was made by the L'Aquila theoretical group, F~Villante et al. ('11); also 
PhD dissertations by S~Marcocci, I~Drachnev, X~Ding and D~Guffanti ('16-'19) deserve praise\footnote{Exciting results on geo-neutrino were achieved in KamLAND and again in Borexino with the help of F~Mantovani, M~Lissia, G~Fiorentini ('03) .}.
\paragraph{\em BBN confronts CMB:}
An improved description of deuterium dynamics in early universe conditions was obtained by LUNA.
This allowed the refined BBN simulations of the PArthENoPE team of theorists,  who continue to develop Gamow's vision,  to affirm the excellent agreement of the BBN predictions and the baryonic mass fraction  $\Omega_b h^2$ measured from the CMB studies. 
Moreover this has implications on  the `effective number' of neutrinos, which turn out to agree with the standard model expectation. 
\paragraph{\em One remark on SN1987A:}
Supernovae   are discussed by C~Volpe. Let me highlight  just one result that illustrates once again and very clearly the importance of interdisciplinarity. 
Probably the hottest issue on SN1987A, until recently, was the absence of an identifiable neutron star. But thanks to 2015 astrophysical simulations by Miceli, Orlando et al.~the first indications of such a star were found: Observational support  in the X-ray and infrared bands was obtained. Needless to say, this result is very important for many disciplines. 
\paragraph{\em High-energy neutrinos:}
Observatories for high-energy neutrinos and $\gamma$-rays, produced as secondary radiation by CR, were conceived in the late 1950s. A dozen years ago
Icecube results finally revealed a signal attributable to cosmic neutrinos of this type, even w/o arriving at a clear understanding of what the sources are.
New tests are needed and they will carried out in Km3NET, GVD and elsewhere with much more precise pointing and ability to investigate the Milky Way accurately. These experiments will explore extreme    environments, high pressures,  allowing new science marine, glaciology, geophysics…

\section{Doubtful cases}
In this section I focus on a few influential scientific works (some published, some not) that offer me opportunities to advance critical comments on the topics  under discussion.
\paragraph{\em Neutrinoless double beta decay:} Many papers have pointed out that the observation of 
neutrinoless double beta decay does not imply Majorana neutrinos, in particular \cite{bb}. 
However,  J~Schechter and JWF~Valle in
{\em Neutrinoless double-$\beta$ decay in SU(2)$\times$U(1) theories,}  Phys.~Rev.~D, 22, 2227 (1981) claim the contrary already from the abstract,
appealing to a ``natural'' theory. This work is very famous; e.g., Wikipedia talks of ``Schechter-Valle theorem''. Note that
1)~the term `theorem’ does not appear in the paper;
2)~none of the previous works are cited;
3)~it all boils down to the definition of `natural' theory, that in the paper reads
``no special adjustment of parameters''.
Even if the claim is reasonable, the sense of confidence given by the word `theorem' is not only unjustified but also dangerous, as it risks having misleading effects. In fact, if the implication works  
`Majorana neutrino mass $\Rightarrow$ $0\nu2\beta$,' i.e.~backwards,  
 we should work harder to understand the mass of neutrinos, instead of focusing all our efforts only on the experimental search for the transition.
 
\paragraph{\em Superluminal neutrinos:} 
Today, the case of superluminal neutrinos does not raise  interpretative doubts. Strictly speaking, the problem reduced to a preprint with a biased measurement, corrected before publication; but this does not mean that the case does not deserve reflection. In fact, one very prestigious journal, Nature, decided to present the lead author of the analysis among the most influential scientists of 2011. Moreover, Glashow \& Coleman had outlined  the possibility that any particle, including neutrinos,  have their own limiting velocity (1998). These are different circumstances, but they provide us with an opportunity to reason about the distortions of the concept of innovation: 
The public's desire for sensational results and the eagerness of some physicists to speculate on remote possibilities may adversely affect neutrino physics, which normally develops in small steps and is more often fuelled by patience than by hasty excitement.

\paragraph{\em Experimental  search for light sterile neutrinos:} The last  case stems from  an (unpublished) but very influential work of 2012, titled
{\em Light Sterile Neutrinos: a White Paper.}
For non-English speaking people like me, the term `White paper' is striking\footnote{The definitions 
from the Merriam-Webster read ``a government report'' or ``a detailed or authoritative report''.} and 
the impression becomes even stronger when scrolling through the list of authors, which contains many of the colleagues I value most. 
However, reading the work, I found myself strongly puzzled by a certain propensity to `push' the field.
In my opinion, it is very different to evaluate scientific cases and then carefully consider whether and how to carry out an experiment, or vice versa, to decide to carry out experiments and work on building custom scientific cases.

\section{Two tips from the past and points for discussion}
I would like to start with some advice from two founders of neutrino science on the points we are interested in. The first concerns innovation and is taken from the review work of Bruno Pontecorvo~\cite{rr}
\begin{quote}
{\footnotesize ``one should neither underestimate the importance of high-energy neutrino physics\footnote{In the parlance of Pontecorvo, the term 
`high-energy neutrino physics' means what we call today `accelerator neutrino physics',
including the long-baseline experiments.}, nor overestimate it. This is not pessimism, but an appeal to avoid routine''

}
\end{quote}
The second  is by John Bahcall and concerns interdisciplinarity:
\begin{quote}
{\footnotesize   ``everyone agrees to do interdisciplinary science but no one wants the money to come from their budget''}
\end{quote}
but 
unfortunately it relies only on my memory and I cannot guarantee for accuracy\footnote{However, recall that Bahcall participated in the evaluation of important US scientific projects.}.
Let us summarize the discussion.\\[1ex]
\noindent {\bf\LARGE N}eutrino physics has always been interdisciplinary - and still is today. It should be added that the data do not speak for themselves; e.g., concepts such as ‘multi-messenger’ help, but on their own are not enough to continue doing well;\\
{\bf\LARGE O}ccurred successes are due to the synergy between good theories and experiments. The main driving force seems not to have been an abstract aspiration for innovation, but rather a consistent commitment to scientific goals and consequently an aptitude for careful planning. Technology matters, but not exclusively;\\
{\bf\LARGE W}e are invited by history of neutrino physics to renew the connections between physics, mathematics, astrophysics, cosmology - as if to say, to cross portals.

{\footnotesize 
\section*{Acknowledgments}
I am grateful for useful discussions to S~Lavignac, M~Miceli, A~Papa, P~Sapienza \& ZZ~Xing. 
Talk based on the last two works in~\cite{rr} and on~\cite{vv} where all references to the various aspects touched upon can be found. 
With support of grant  2022E2J4RK {\em PANTHEON: Perspectives in Astroparticle and Neutrino THEory with Old and New messengers,} PRIN 2022, funded by the Italian Ministero dell’Universit\`a e Ricerca (MUR) \& European Union – Next Generation EU.

{\footnotesize
\appendix
\section{Epistemological notes}

An attitude of empiricism can be summed up with the phrase {``If it cannot be measured, it does not exist.''} Although it has often proven to be a valid approach, it is not without risks. For example, it leads one to believe that it is better to focus on models that allow something to be measured, which could be really misleading. It is not difficult to feign in our minds scenarios  formally consistent, but this does not mean that they all have the same value, or even that they have one. On the contrary, there are theories based on profound principles that take time to explore thoroughly.

Curiously enough, the opposite attitude is not infrequently found in some colleagues allegedly interested in `physics beyond the standard model’, who nonetheless downplay, or sometimes even choose to ignore, the only evidence of physics beyond the standard model that is based on experiments and that may allow further progress - I mean neutrino masses\footnote{E.g., P~Woit, whom I usually appreciate, in the  lecture  
{\em The Forgotten Geometry: A New Path to Unification} 
stated that:  ``the first version of this theory didn't have masses for the neutrinos, but it turns out you can throw in some right-handed neutrino fields, and it all  works exactly, you know, as you expect so far," 
that I consider a hasty  summary.}. This attitude is just as if not more dangerous than the previous one:
It leads to an underestimation of the importance of ongoing enterprises and prevents us from understanding where neutrino science can realistically go.

}
}
\end{document}